\documentclass[12pt]{article}

\usepackage{amsmath}
\usepackage{graphicx}

\newcommand{\eq}{\begin{equation}}
\newcommand{\eqx}{\end{equation}}
\newcommand{\eqs}{\begin{equation*}}
\newcommand{\eqsx}{\end{equation*}}
\newcommand{\eqn}{\begin{eqnarray}}
\newcommand{\eqnx}{\end{eqnarray}}
\newcommand{\alg}{\begin{align}}
\newcommand{\algx}{\end{align}}

\newcommand{\f}[2]{\frac{#1}{#2}}

\newcommand{\lm}{\lambda}

\newcommand{\om}{\omega}
\newcommand{\gm}{\gamma}
\newcommand{\qqqq}{\quad\quad\quad\quad}

\newcommand{\slii}{{\tt sl(2)}\ }

\newcommand{\tr}{\mbox{\rm tr}\,}
\newcommand{\nn}{{\cal N}}
\newcommand{\oo}[1]{{\cal O}\left(#1\right)}

\title{Four loop twist two, BFKL, wrapping and strings}

\author{Zoltan Bajnok$^{a}$\thanks{e-mail: {\tt bajnok@elte.hu}},\ \  
Romuald A. Janik$^{b}$\thanks{e-mail: {\tt ufrjanik@if.uj.edu.pl}}\\ and
Tomasz {\L}ukowski$^b$\thanks{e-mail: {\tt tomaszlukowski@gmail.com}} \\ \\  
${}^a$ Theoretical Physics Research Group \\
       Hungarian Academy of Sciences \\
       1117 Budapest, P\'azm\'any s. 1/A \\
       Hungary\\
${}^b$ Institute of Physics\\
Jagellonian University,\\
ul. Reymonta 4, \\
30-059 Krak{\'o}w\\
Poland}

\begin{document}

\maketitle

\begin{abstract}
The anomalous dimensions of twist two operators have to satisfy certain
consistency requirements derived from BFKL. For $\nn=4$ SYM it was
shown that at four loops, the anomalous dimensions derived from the
all-loop asymptotic Bethe
ansatz do not pass this test. In this paper we obtain the remaining
wrapping part of these anomalous dimensions from string theory and
show that these contributions exactly cure the problem and lead to
agreement with both LO and NLO BFKL expectations.
\end{abstract}

\vfill 

\section{Introduction}

The integrable structures which appear both on the gauge theory side
\cite{Minahan:2002ve,Beisert:2003tq,Beisert:2003yb,kor1,Dolan:2003uh} and the
string theory side \cite{Bena:2003wd} of the AdS/CFT correspondence
\cite{adscft} give hope to find the full spectrum of both theories. On the
gauge theory side we are interested in finding the anomalous
dimensions of gauge  
invariant operators in $\nn=4$ super Yang-Mills theory, while on the
string theory 
side we would like to determine the energies of the superstring
excitations in  
the $AdS_5\times S^5$ background. 

A lot of progress has been done for specific type of operators, namely
with large numbers of fields, and for the corresponding infinitely long/fast
strings (strings with large charges). The S-matrix for elementary
excitations has been 
predicted from the symmetry algebra up to the overall scalar factor \cite{B}
which was finally fixed in \cite{BHL,BES} using the crossing symmetry
\cite{CROSS}. The Bethe Ansatz Equations have been derived
\cite{Beisert:2003yb,KMMZ,K1,K2,BDS,AFS,S,BS} giving the spectrum of states with large
quantum numbers.

Nevertheless, it is known that the Bethe Ansatz is not valid
for short operators and strings with small charges. On the gauge
theory side it is 
due to the fact that at the order of $\lambda^L$, where $L$ is the length
of the gauge 
invariant operator, wrapping corrections start to play a role. In an 
analogous 
manner \cite{AJK}, on the string theory side virtual corrections
around the worldsheet 
cylinder appear. 
In \cite{JL} the
leading exponential finite size effects at strong coupling for a
single giant magnon were computed from the  
identification with L{\"u}scher-like corrections (see also
\cite{HS,SSV,HJL}) and turned out to agree exactly with the expression
obtained from classical string solution \cite{AFZmagnon}. 
At weak coupling, wrapping corrections were analyzed from the gauge
theory perturbative point of view in \cite{Sieg}.
The wrapping contribution to the Konishi operator anomalous
dimension at four loops has been explicitly computed within
perturbative gauge theory 
\cite{FSSZ} (a subsequent independent ab-initio perturbative computation of
\cite{Velizhanin08} which included both the wrapping and non-wrapping
graphs reaffirmed this result). The same result was
obtained purely within string theory from the $AdS_5\times S^5$ string
sigma model in \cite{BJ08}.   

The wrapping effects appear in the most manifest way for the shortest possible
operator. The simplest non protected operators, which are 
generalizations of the Konishi operator, are twist two  operators: $\tr
ZD^{2M} Z$. Apart from being a natural testing ground for
wrapping effects, which should appear at 4 loops, their anomalous
dimensions are also interesting for 
their own sake as they are intertwined with various seemingly
unrelated physical processes and have rich analytical properties. Moreover
these operators (in fact their very close analogues) are also of big
importance in ordinary QCD (see in particular
\cite{Jaroszewicz:1982gr} in the context of the present paper). 

Their anomalous dimensions have been computed perturbatively (in QCD)
up to 3 loops \cite{MVV} and using the maximum transcendentality
conjecture \cite{Kotikov:2002ab} for $\nn=4$ SYM
the corresponding answer in $\nn=4$ SYM has been extracted
\cite{Kotikov:2003fb,Kotikov:2004er,Kotikov:2006ts}. This has been
found to agree with the prediction from the asymptotic Bethe
Ansatz. Subsequently the Bethe Ansatz answer for four loops was
computed \cite{KLRSV} and found to satisfy the maximal transcendentality
principle. It was, however, demonstrated that the Bethe Ansatz answer
is in conflict with predictions  from the BFKL equation
\cite{Lipatov:1976zz}. It was 
pointed out that the missing part of the four-loop answer should come from 
wrapping effects. The motivation of our paper is to fill this gap and find
the leading wrapping correction for general twist two operators and
reexamine the consistency with BFKL.   

Thus our aim is to extend the calculations of the wrapping corrections to
the case of arbitrary twist two operators. We will adapt the method
used in the 
case of Konishi operator in \cite{BJ08}, based on the integrability properties
of the worldsheet quantum field theory of the superstrings in $AdS_5\times
S^5$. The leading contribution from wrapping will appear at order $\lambda^4$
which exactly matches the predictions coming from TBA type
considerations \cite{AJK}.    

The operators considered in this paper are part of a larger family of
interesting operators from the \slii sector made up from $M$
derivatives (spin) and $J$ complex scalars (twist): $\tr(D^{s_1}Z \dots
D^{s_J}Z)$. For high
er twist,
there is also a wide range of interesting phenomena e.g. the large spin
($M$) and twist ($J$) behaviour of their anomalous 
dimension $\Delta (M,J)$  is determined by the BES/FRS integral equations
\cite{BES,FRS}. Remarkably, in the limit when both $M,J\to \infty$ such that
$j=\frac{J}{\log M} $ is kept fixed the leading logarithmic behaviour 
is governed by the cusp anomalous dimension and the ground-state energy 
density of the O(6) $\sigma $ model 
\cite{AM,Casteill,BKK,Fior1,BK,Fior,Fior2,G,RT,BBBKP}. We
expect that the methods presented in the present paper should also be
applicable for the higher twist case.

The plan of this paper is as follows. In section 2 we briefly introduce
the twist two operators and their main properties. In section 3 the
formula for 
wrapping correction for general twist two operator is given. In section 4 we
point out all ingredients needed to compute the leading wrapping
correction. All relevant calculations ispresented in section 5, while the
final answer is obtained in section 6. Section 7 shows that the
wrapping corrections do not change the large $M$ asymptotic behaviour of
anomalous dimension for twist two operators, consequently, the cusp
anomalous dimension (`scaling function') is untouched by 
wrapping. In section 8 we analytically continue the final result to
$M=-1+\omega$ and then compare it with the BFKL prediction. We close the paper
with a discussion and several appendices. 

\section{Twist two operators and BFKL}

Twist two operators in $\nn=4$ SYM are operators which are made from
two $Z$ fields ($J=2$) and an arbitrary (but even) number of
derivatives $M$ in a 
fixed light cone direction. Schematically these operators are thus of
the form $\tr ZD^{M} Z$. For each even $M$ there is a unique highest 
weight non BPS operator and so we may define the anomalous dimension
\eq
\label{e.deltatwist}
\Delta(M,J=2)=2+M+\sum_{l=1}^\infty \gm_{2l}(M) g^{2l}
\eqx
where $g^2=\lm/16\pi^2$. Here the index $l$ denotes the
perturbative loop order. This quantity can be computed from the
asymptotic Bethe ansatz in the \slii sector exactly up to 3 loops. The
answer from the Bethe ansatz at 4 loops and higher will have to be
supplanted by the contribution of so-called `wrapping
interactions'. The aim of this paper is to compute this contribution
at 4 loop order from the string sigma model in $AdS_5 \times
S^5$. This arises due to the identification of the anomalous
dimensions with energies of string states in $AdS_5 \times S^5$. Since
the worldsheet QFT is integrable, and we know the exact S-matrix, we
know the full on-shell data of the worldsheet QFT at infinite
volume. We may therefore study the spectrum of energies around the
infinite volume limit. The leading piece is contained in the Bethe Ansatz
(identical to the asymptotic Bethe ansatz), while the leading virtual
corrections computed from generalized multiparticle L{\"u}scher
formulas provide a way to compute {\em exactly} the 4-loop wrapping
corrections. Thus we may split the 4-loop coefficient of
(\ref{e.deltatwist}) into
\eq
\gm_8(M)=\gm_8^{Bethe}(M)+\gm_8^{wrapping}(M)
\eqx
The Bethe ansatz term has been computed in \cite{KLRSV} and can be
found in table~1 of that reference. The aim of this paper is to
compute the wrapping part $\gm_8^{wrapping}(M)$ from multiparticle L{\"u}scher
formulas for the $AdS_5\times S^5$ worldsheet QFT. This is a
generalization of the computation of the 4-loop anomalous dimension of
the Konishi operator \cite{BJ08} which corresponds to
\eq
\gm_8^{wrapping}(2)=324+864\zeta(3)-1440 \zeta(5)
\eqx

The function $\Delta(M,J)$ has numerous
interesting properties. Firstly, its large $M$ limit is related to the
cusp anomalous dimension \cite{Korchemsky:1988si}:
\eq
\lim_{M\to \infty} \Delta(M,J)-J-M \sim 2\gm_{cusp}(g) \log M
\eqx   
It is therefore expected that the coefficient of $\log M$ does not
depend on $J$. The cusp 
anomalous dimension can be investigated both from the perturbative
side \cite{Bern:2006ew} and from the
strong coupling side 
\cite{Benna:2006nd,Alday:2007qf,BKK,Kostov:2008ax,GKP,Frolov:2002av,Roiban:2007jf} with an 
interpolating answer following from the BES equation derived from the
asymptotic Bethe ansatz \cite{BES}. The applicability of the Bethe
ansatz analysis rests on the independence of $\gm_{cusp}(g)$ on $J$ so
that wrapping interactions do not contribute in this limit. This can
be argued both on the perturbative side \cite{korgorsky} and on the
strong coupling string side \cite{tseytlinkrucz}, but it would be 
interesting to
verify directly that wrapping contributions will vanish in this
limit. This is among the aims of the present paper where we
verify this for $J=2$ and 4-loop order. 
Once these wrapping corrections are obtained, thus completing our
knowledge of 4 loop anomalous dimensions of twist two operators, there
are also other very interesting features of their large $M$ limit
which could be investigated in particular its `reciprocity' properties
\cite{bassokor}, further studied in \cite{reciprocity}.

Secondly, the coefficients of (\ref{e.deltatwist}) at $l$-loop,
$\gm_{2l}(M)$ are expected to obey, for $\nn=4$ SYM, the principle of
maximum transcendentality \cite{Kotikov:2002ab}. This means that they
are expressed in 
terms of (nested) harmonic sums and $\zeta$ functions such that the
degree of transcendentality of $\gm_{2l}(M)$ equals $2M-1$. The degree
of transcendentality for a product is defined to be the sum of the
degrees of each factor. The degree of transcendentality of $\zeta(n)$
is defined to be $n$ while the
degree of the harmonic sum $S_{i_1,i_2,\ldots, i_k}(M)$ is $\sum_i
|i_k|$. The harmonic sum $S_{i_1,i_2,\ldots, i_k}(M)$ is defined
recursively as
\eq
S_{i_1,i_2,\ldots, i_k}(M)=\sum_{n=1}^M \f{{\rm sign}(i_1)^n}{n^{|i_1|}}
S_{i_2,\ldots, i_k}(n) 
\eqx
while the elementary harmonic sums with a single index are given by
\eq
S_{j}(M)= \sum_{n=1}^M \f{{\rm sign}(j)^n}{n^{|j|}}
\eqx
In \cite{KLRSV} it was verified that the part of the 4-loop result
following from the asymptotic Bethe ansatz indeed is composed of terms
of transcendentality degree 7. We would like to verify in the present
paper that the contribution of wrapping corrections will also have the
same degree of transcendentality.

The third and perhaps the most nontrivial property of the anomalous
dimensions (\ref{e.deltatwist}) is the interrelation with BFKL
\cite{Lipatov:1976zz} 
and NLO BFKL \cite{NLOBFKL} equations which describe perturbatively
gauge theory high energy 
scattering in the Regge limit. From this formalism one
obtains a specific prescription for the pole structure of the {\em
  analytical continuation} of $\Delta(M,J=2)$ to $M=-1+\om$. The
$l$-loop coefficients of (\ref{e.deltatwist}) have to have the
following pole structure around $M=-1+\om$ \cite{KLRSV}:
\eqn
\gm_{2}(\om) &\sim& -4\left( \f{2}{\om} + 0 +\oo{\om} \right) \\
\gm_{4}(\om) &\sim& -16\left( \f{0}{\om^2} + \f{0}{\om} +\oo{\om^0} \right) \\
\gm_{6}(\om) &\sim& -64\left( \f{0}{\om^3} + \f{\zeta(3)}{\om^2} +
\oo{\f{1}{\om}} \right) \\ 
\gm_{8}(\om) &\sim& -256\left( \f{4\zeta(3)}{\om^4} +
\f{\f{5}{4}\zeta(4)}{\om^3}  +\oo{\f{1}{\om^2}}  \right)
\eqnx
where the leading pole comes from the BFKL equation while the second
one is a consequence of NLO BFKL. It has been verified \cite{KLRSV} that the
1-,2- and 3-loop result exactly agrees with the above BFKL and NLO
BFKL predictions. The main conclusion of \cite{KLRSV} was that the Bethe
ansatz part of $\gm_{8}(\om)$ does not satisfy this constraint. Its
expansion around $M=-1+\om$ is\footnote{We are grateful to
  Vitaly Velizhanin and the authors of \cite{KLRSV} for informing us of
  this explicit expansion.}  
\eq
\label{e.bethean}
\gm_{8}^{Bethe}(\om) \sim 256\left( \f{-2}{\om^7}+ \f{0}{\om^6} +
\f{8\zeta(2)}{\om^5} -\f{13\zeta(3)}{\om^4} -\f{16 \zeta(4)}{\om^3}
+\oo{\f{1}{\om^2}} \right)
\eqx
The main motivation of this work is to compute the corresponding
wrapping contribution $\gm_{8}^{wrapping}(M)$ and to check whether the
sum $\gm_8(\om)=\gm_8^{Bethe}(\om)+\gm_8^{wrapping}(\om)$ has the
correct analytical properties required by LO and NLO BFKL. 

\vfill
\pagebreak

\section{Wrapping correction for twist two operators}

\begin{figure}[h]
\begin{center}
\includegraphics[width=5cm]{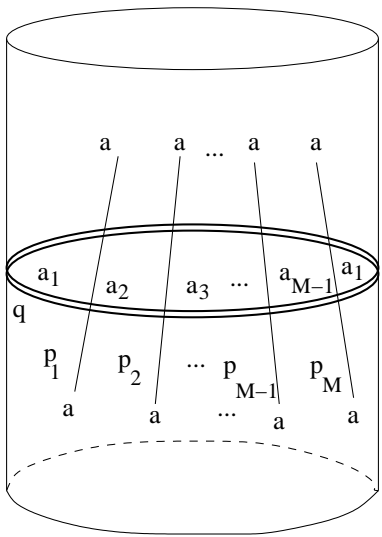}
\end{center}
\caption{Multiparticle L{\"u}scher correction. The vertical lines
  represent the physical particles forming the multiparticle state,
  while the double line loop represents the on-shell `virtual'
  particle with complex momentum.}
\end{figure}

The aim of this section is to extend the calculation of the leading
wrapping correction of the Konishi operator to generic twist two operators.
This amounts to calculating the leading finite size energy correction
of a specific $M$ particle state. In \cite{BJ08} the authors conjectured
a L\"uscher type formula for generic multiparticle states and successfully
applied it for the Konishi case: $M=2$. The formula consists of two
parts: The first describes the effect how the finite volume modifies
the particles' quantization conditions and shifts their momenta, while
the second is due to virtual particles propagating around the cylinder
and directly changes the energy as
\begin{equation}
\Delta
E(L)=-\sum_{Q=1}^\infty \int_{-\infty}^{\infty}\frac{dq}{2\pi}
\mbox{STr}_{a_{1}} 
\left[S_{a_{1}a}^{a_{2}a}(q,p_{1})S_{a_{2}a}^{a_{3}a}(q,p_{2})\dots 
  S_{a_{M}a}^{a_{1}a}(q,p_{M})\right]
e^{-\tilde{\epsilon}_{a_{1}}(q)L}\label{eq:Luscher}
\end{equation}  
The content of this formula is schematically represented in Fig.~1. 
This formula applies for an $M$ particle state with all particles
of type $a$, such that their consecutive self-scatterings preserve
this state and determine their momenta $p_{i}$ by the BA equations.
The matrix $S_{ba}^{ca}(q,p)$ describes how a virtual particle of
type $b$ with momentum $q$ scatters on the real particle of type
$a$ and momentum $p$. The exponential factor can be interpreted
as the propagator of the virtual particle. What makes difficult to
apply this formula in practice is that we have to sum over all possible
virtual particles in the theory (both fundamental magnons $Q=1$ and the
infinite tower of their bound states $Q>1$), their polarizations $a_1$ and,
even more, over all possible intermediate states $a_{2},\dots,a_{M}$.

Let us focus on the leading wrapping correction for the anomalous
dimension of a twist two operator. Since the modification of the BA
equation appears at order $g^{8}$ its contribution to the energy
will be subleading and it is sufficient to analyze equation
(\ref{eq:Luscher}). 
In the next section we explain all the ingredients of this equation
specified to the twist two case, while in the subsequent one we perform
the actual calculation.

\section{Main ingredients of the wrapping correction}

In order to apply formula (\ref{eq:Luscher}) for twist two operators
we have to explain the following ingredients: What is the state such
an operator corresponds to and how are their momenta determined by
the BA equation? What are the virtual excitations, what is their exponential
damping factor and how they scatter on the state that corresponds
to the twist two operator? Let us investigate these questions in this
order.

\subsection*{Multiparticle states corresponding to twist two operators}

The state which corresponds to twist two operators is an $M$ particle
state with $M$ even. According to the $su(2\vert2)\otimes su(2\vert2)$
classification each particle 
belong to the \slii sector of the fundamental representation,
which is realized in terms of the totally anti-symmetric representation,
thus  has label $a=(1,\dot{1})$%
\footnote{We use the convention in which fermionic coordinates have
  labels $1,2$ 
while bosonic ones $3,4$. See \cite{BJ08} for details. %
}. The volume is $L=2$ and the momenta of the particles are determined
by the BA equation. At leading order in $g$ their rapidities \[
u(p)=\frac{1}{2}\cot\frac{p}{2}\sqrt{1+16g^{2}\sin^{2}\frac{p}{2}}\]
are given as the roots of Baxter's $Q$ function, $P_M(u)$. This is
a polynomial of order $M$ which is given explicitly (at 1-loop
  level\footnote{This expression has been generalized to higher loops
  in \cite{Kotikov:2008pv}. However we will not need these expressions
  for our computation as they would contribute only at higher loop
  orders.}) by the generalized hypergeometric function (see
  \cite{Derkachov:2002tf,ES} 
and the methods of \cite{FK})
\[
P_M(u)=\, _{3} F_{2}(-M,M+1,\frac{1}{2}-iu;1,1\vert1)
\]
Since this polynomial is even each rapidity $u_{k}$ comes in pairs:
\begin{equation}
P_M(u)\propto \prod_{k=1}^{M}(u-u_{k})=(u-u_{1})(u+u_{1})
\dots(u-u_{\frac{M}{2}})(u+u_{\frac{M}{2}})\label{eq:BaxtersQ}
\end{equation} 
This is a useful form that we will use later on.

\subsection*{Virtual particles and the exponential factor}

The virtual particles turn out to belong to the completely 
anti-symmetric representation
of $su(2\vert2)\otimes su(2\vert2)$ first discussed in
\cite{AFbound}. 
This representation exists for
any integer 
$Q$ and has dimension $(4Q)^{2}$. The dispersion relation of their
particles leads at leading order to the exponential factor \[
e^{-\tilde{\epsilon}_{Q}(q)L}=e^{-2L\mathrm{arcsinh}
  \frac{\sqrt{q^{2}+Q^{2}}}{4g}}=  
\left(\frac{z^{-}}{z^{+}}\right)^{L}=\frac{4^{L}g^{2L}}{(q^{2}+Q^{2})^{L}}\]  
where here and later on we use that \[
z^{\pm}(q)=\frac{q+iQ}{4g}\left(\pm1+\sqrt{1+\frac{16g^{2}}{q^{2}+Q^{2}}}
\right)\]
Observe that only $z^{+}$ scales like $g^{-1}$ but $z^{-}$ goes
as $g$ in the weakly coupled regime.

\subsection*{The scattering matrix}

In \cite{BJ08} it was described how a particle with charge $Q$ and
parameters $z^{\pm}$ scatters on a fundamental particle with $Q=1$
and parameters $x^{\pm}$ \[
x^{\pm}(u)=\frac{2u\pm i}{4g}\left(1+\sqrt{1-\frac{16g^{2}}{(2u\pm
    i)^{2}}} \right)\]
The scattering matrix can be determined, using the superfield methods
of \cite{AFSmatrix}, from the requirement that
it commutes with the symmetry charges of $su(2\vert2)\otimes
su(2\vert2)$ up to a 
scalar factor which was obtained from the bootstrap. Let us recall
the scalar and matrix part of the scattering matrix.

\subsubsection*{Scalar part}

In the bootstrap procedure the composite particle $z^{\pm}$ is realized
in terms of its individual constituents $z_{i}^{\pm}$ such that the
bound-state condition is realized: $(z^{-}=z_{1}^{-},z_{1}^{+})$,
$(z_{2}^{-}=z_{1}^{+},z_{2}^{+}=z_{3}^{-}),\dots, 
(z_{Q}^{-}=z_{Q-1}^{+},z_{Q}^{+}=z^{+})$.
The scalar factor of the $Q-1$ scattering is then nothing but the
product of the individual scalar factors of the $1-1$ scatterings:
\begin{equation}
S_{Q-1}^{sl(2)}(z^{\pm},x^{\pm})=\prod_{i=1}^{Q}S_{1-1}^{sl(2)}(z_{i}^{\pm},
x^{\pm}) 
\label{e.sl2fuse}\end{equation}
where \begin{equation}
S_{1-1}^{sl(2)}(z^{\pm},x^{\pm})=\frac{z^{-}-x^{+}}{z^{+}-x^{-}}
\frac{1-\frac{1}{z^{+}x^{-}}}{1-\frac{1}{z^{-}x^{+}}}\end{equation}
The choice for the constituents is not unique and such configuration
was adopted which had the most $z_{i}^{\pm}$ parameters of order
$g^{-1}$. This leads to the result which, in terms of $q$ and $u$,
can be written as: 

\[
S_{Q-1}^{sl(2)}(q,u)=\frac{(q-i(Q-1)-2u)(2u+i)^{2}}{(q-i(Q+1)-2u)
  (q+i(Q-1)-2u)(q+i(Q+1)-2u)}\]

\subsubsection*{Matrix part}

In describing the matrix part we can exploit the fact that it is a
tensor product of two copies of the same S-matrices. This reflects
the $su(2\vert2)\otimes su(2\vert2)$ nature of the symmetry. Thus
it is enough to analyze one copy of the anti-symmetric representations.
It has $4Q$ states: $2Q$ bosons and $2Q$ fermions. From the
$su(2\vert2)$ symmetry 
it is possible to calculate how these particles scatter on the particles
of the fundamental representation, see \cite{BJ08} for the result.

For calculating the wrapping correction we need only the
$S_{1I}^{1J},\, I,J=1,\dots4Q$ 
part of the matrix. We will interpret it as a matrix acting on the
anti-symmetric representation of charge $Q$ and denote it by
$S_{Qmatrix}^{sl(2)}(u,q)$. 
Lets us go through systematically all possible matrix elements. We
use the notation of \cite{BJ08} adapted to anti-symmetric representations.
The first bosonic state $I=1$ can scatter into itself \[
SB0(q,u):=S_{11}^{11}(q,u)\]
The bosonic state $I=j$, $j=2,\dots,Q$ can scatter either into itself,
or into the other bosonic state $J=j+Q$. In a similar manner this
other bosonic state with $J=j+Q$ can either scatter into itself or
scatter back to $I=j$. Thus for each $j=2,\dots,Q$ these scatterings
form a $2\times2$ matrix \[
SB(q,u,j+1)=\left(\begin{array}{cc}
S_{1\, j}^{1\, j}(q,u,j) & S_{1\, j}^{1\, j+Q}(q,u,j)\\
S_{1\, j+Q}^{1\, j}(q,u,j) & S_{1\, j+Q}^{1\, j+Q}(q,u,j)\end{array}\right)\]
The boson with index $I=Q+1$ scatters again into itself
\[
SBQ(q,u)=S_{1\, Q+1}^{1\, Q+1}(q,u)\]
The fermionic part of the scattering is much simpler, since it is
diagonal and is the same for both fermionic particles:
\[
SF(q,u,j+1)=S_{1\,2Q+j}^{1\,2Q+j}(q,u)=S_{1\,3Q+j}^{1\,3Q+j}(q,u)\]
The explicit form of these matrix elements can be found in the Appendix
of \cite{BJ08}. Here, for convenience, we grouped them slightly differently
by shifting some of the labels. (For the paper to be self contained,
we list the needed  
matrix elements explicitly in Appendix D).

\section{The calculation of the wrapping correction}

Having introduced all the needed quantities we can formulate the wrapping
correction as
\[
\Delta E=-\sum_{Q=1}^{\infty}\int_{-\infty}^{\infty}\frac{dq}{2\pi}
\prod_{k=1}^{M}S_{Q-1}^{sl(2)}(q,u_{k})\Bigl(\mbox{STr}(\prod_{k=1}^{M}
S_{Qmatrix}^{sl(2)}(q,u_{k}))\Bigr)^{2}\frac{16g^{4}}{(q^{2}+Q^{2})^{2}}\]
Where the super-trace part can be further elaborated as \begin{eqnarray}
\mbox{STr}(\prod_{k=1}^{M}S_{Qmatrix}^{sl(2)}(q,u_{k})) & 
= & \prod_{k=1}^{M}SB0(q,u_{k})+\prod_{k=1}^{M}SBQ(q,u_{k})\nonumber \\
 &
& \hspace{-2cm}-2\sum_{j=0}^{Q-1}\prod_{k=1}^{M}SF(q,u_{k},j)+
\sum_{j=1}^{Q-1} 
\mbox{Tr}\left(\prod_{k=1}^{M}SB(q,u_{k},j)\right)
\label{eq:STr}
\end{eqnarray}
We evaluate this part carefully in Appendix A and here we just summarize
the main steps. Since we expect the wrapping contribution to appear
at order $g^{8}$, and there is an explicit $g^4$ dependence coming
from the `exponential' part, the super-trace part has to 
be of order $g^2$, so its leading term should vanish.
We check this requirement first and then make a systematical small
$g^{2}$ expansion of each term later. Schematically we expand each
function as $f(g)=f_{0}(1+g^{2}\delta f+O(g^{4}))$. The key feature
why we were able to do the calculation is, that the leading order
part of the bosonic matrices $SB(q,u,_{k},j)_{0}$ can be diagonalized
in a $u_{k}$-independent way. This enables us to rewrite the bosonic
matrix contribution as a sum over the contributions of the two eigenvalues
$SB^{\pm}(q,u_{k},j)_{0}$. Explicitly we found that \[
SB^{+}(q,u_{k},j)_{0}=\frac{i+2ij+q-iQ-2u}{i+q-iQ-2u}\]
Additionally, when we extended the summation for these eigenvalues
(from $1$ to $0$ for $+$ and form $Q-1$ to $Q$ for $-$) we could
incorporate the contributions of the two separate bosonic terms
$SB0(q,u_{k})_{0}$ 
and $SBQ(q,u_{k})_{0}$. We also observed that
$SB^{-}(q,u_{k},j+1)=SB^{+}(q,u_{k},j)\frac{2u+i}{2u-i}$ 
such that the $-$ bosonic summation can be also shifted and the zero
order part for wrapping has the form: 
\[
\mbox{STr}\Bigl(\prod_{k=1}^{M}S_{Qmatrix}^{sl(2)}(q,u_{k})\Bigr)_{0}=
-2\sum_{j=0}^{Q-1}\prod_{k=1}^{M}SF(q,u_{k},j)_{0}+
2\sum_{j=0}^{Q-1}\prod_{k=1}^{M}SB^{+}(q,u_{k},j)_{0} 
\]
Further checking that the fermionic parts and the bosonic parts are
the same for each $j$,
($SF(q,u_{k},j)_{0}=SB^{+}(q,u_{k},j)\sqrt{\frac{2u+i}{2u-i}}$), 
we could see that they completely annihilate each other at this zeroth
order. 

In doing the calculation at first order we performed the same steps: We
diagonalized the bosonic contributions upto $O(g^{2})$, we extended
the bosonic summation to incorporate the extra separate bosonic pieces,
shifted the summation for the second bosonic part, and finally exploited
the fact that the fermionic and bosonic zero order terms are the same
for any $j$. As a result we arrived at formula 
\begin{eqnarray*}
\mbox{STr}\Bigl(\prod_{k=1}^{M}S_{Qmatrix}^{sl(2)}(q,u_{k})\Bigr) & =
 & g^{2}\sum_{j=0}^{Q-1}\Bigl(\prod_{k=1}^{M}SB^{+}(q,u_{k},j)_{0}\Bigr)
\sum_{k=1}^{M}\delta SBF(q,u_{k},j)
\end{eqnarray*}
where after some lengthy but straightforward calculations we found
that 
\[
\delta SBF(q,u_{k},j)=\frac{16}{1+4u_{k}^{2}}\left[\frac{1}{2j-iq-Q}- 
\frac{1}{2(j+1)-iq-Q}\right]
\]
Exploiting the very simple $u_{k}$ dependence of the summand we can
recognize the one loop BA energy and replace it with the harmonic
sum $S_{1}(M)$ as 
\[
\sum_{k=1}^{M}\frac{16}{1+4u_{k}^{2}}=8S_{1}(M)=8\sum_{l=1}^{M}\frac{1}{l}\]
In order to abbreviate future formulas we will suppress the arguments
of harmonic sums if they are $M.$ The equation (\ref{eq:BaxtersQ})
can be used to write the result in a more economical way:
\[
\prod_{k=1}^{M}SB^{+}(q,u_{k},j)_{0}=
\frac{P_M(\frac{1}{2}(q-i(Q-1)+ij))}{P_M(\frac{1}{2}(q-i(Q-1)))}
\]

\subsection*{The final form of the wrapping correction}

We arrived at one of the main results of the paper. We merely have
to collect all the ingredients of the wrapping correction: the exponential
factor, the scalar factor and the matrix part. There is an elegant
way of writing the whole wrapping correction in terms of Baxter's
Q function\footnote{We denote the Baxter's function as $P_M(.)$ in order
to avoid confusion with the $Q$ parameter.} as 
\eq
\label{e.integral}
\Delta E=-64g^{8}\, S_{1}^{2}\,\sum_{Q=1}^{\infty}\int_{-\infty}^{\infty}\frac{dq}{2\pi}
\frac{T_M(q,Q)^{2}}{R_M(q,Q)}\frac{16}{(q^{2}+Q^{2})^{2}}
\eqx
where 
\[
R_M(q,Q)=P_M\Bigl(\frac{1}{2}(q-i(Q-1))\Bigr)P_M\Bigl(\frac{1}{2}(q+i(Q-1))
\Bigr) P_M\Bigl(\frac{1}{2}(q+i(Q+1))\Bigr) P_M\Bigl(\frac{1}{2}(q-i(Q+1))
\Bigr)\] 
and 
\[
T_M(q,Q)=\sum_{j=0}^{Q-1}\left[\frac{1}{2j-iq-Q}-\frac{1}{2(j+1)-iq-Q}\right]
P_M\left(\frac{1}{2}(q-i(Q-1))+ij\right)
\]
An alternative way of writing this function is 
\begin{eqnarray*}
T_M(q,Q) & = & \frac{iP_M\left(\frac{1}{2}(q-i(Q-1))\right)}{q-iQ}
-\frac{iP_M\left(\frac{1}{2}(q+i(Q-1))\right)}{q+iQ}+\widetilde{T}_M(q,Q)
\end{eqnarray*}
where \[
\widetilde{T}_M(q,Q)=\sum_{j=1}^{Q-1}\frac{P_M\left(\frac{1}{2}(q-i(Q-1))+
  ij\right)
-P_M\left(\frac{1}{2}(q-i(Q+1))+ij\right)}{2j-iq-Q}\]
It is not difficult to show that $\widetilde{T}_M(q,Q)$ defines a polynomial.

\subsubsection*{Wrapping correction for odd particle number}

In computing the anomalous dimension of twist two operators the consideration
of odd states played an important role \cite{KLRSV}. This meant an
analytical continuation
of the result from even to odd $M$. The hypergeometric function
determines the allowed rapidities: they are all paired except one
which has $u=0.$ This is not physical since it corresponds to $p=\pi$
and not $p=0$. Nevertheless, we
can try to extend the wrapping correction to this case. The first
problem immediately arises at zeroth order. The extra $p=\pi$ state
gives different extra factors to the two bosonic and fermionic contributions.
Namely the $B^{+}$ part is unchanged, the $B^{-}$ gets an additional
$-1$ and annihilates $B^{+},$ while both fermions got an extra $i$.
Since we would like to maintain for the analytical continuation that
the correction starts at $g^{8}$ 
we suppose that in a theory where $u=0$ is an allowed state the symmetry
between the two fermions are broken and their contributions have opposite
signs. Thus we add extra factors as $i$ and $-i$ to the two fermionic
components. With this assumption we could calculate the wrapping correction
of those odd states. The only difference compared to the even one
resides in the function $T(q,Q)$ which turns out to be 
\[
T_M^{odd}(q,Q)=\sum_{j=0}^{Q-1}\left[\frac{1}{2j-iq-Q}+
  \frac{1}{2(j+1)-iq-Q}\right] 
P_M\Bigl(\frac{1}{2}(q-i(Q-1))+ij\Bigr)
\]
So in essence a `nice' analytical continuation of (\ref{e.integral})
amounts to inserting an additional $(-1)^M$ sign in the definition of
$T_M(q,Q)$.

\subsubsection*{Calculating the integral}

We calculate the integral by closing the contour on the upper half
plane and taking residues over the poles of the integrand. We have
two types of poles. The pole at $q=iQ$ is of kinematical origin since
it does not depend on the scattering matrix. In contrast, the poles
of the function $R(q,Q)$ depends on $u_{k}$ thus are determined
by the dynamics and come from the scattering matrix. They correspond
to $\mu$ terms in the L\"uscher correction. In the paper \cite{BJ08}
an argument, based on kinematical considerations, was given that such
terms are absent in the weakly coupled ($g\to0$) limit, so we expect
their contribution to vanish. Although they do not vanish separately
but when we sum up their contributions for all the bound-states (for
$Q$) the result is indeed zero. We checked this explicitly for the
first few $M$ cases. As a consequence in calculating the integral
for the wrapping correction we can take the residue only at the kinematical
pole $q=iQ$. Subsequently we have to sum the resulting expression for
$Q$ running from $1$ to $\infty$.

\section{Determination of the final form of $\gm_8^{wrapping}(M)$}

Assuming the maximum transcendentality principle, we expect that
$\gm_8^{wrapping}(M)$ has the following structure:
\eq
\label{e.finfirst}
\gm_8^{wrapping}(M)= C_7(M) +C_4(M) \zeta(3)+ C_2(M) \zeta(5)
\eqx
where the coefficients $C_n(M)$ have transcendentality degree $n$.

In Appendix B we have analytically derived the coefficients
of $\zeta(3)$ and $\zeta(5)$ and found that indeed they have the
expected transcendentality structure\footnote{In the following all
  harmonic sums are evaluated at $M$ so we suppress the argument.}:
\eq
C_2(M)=-640 S_1^2  \qqqq C_4(M)=-512 S_1^2 S_{-2}
\eqx
It remains to determine the rational part $C_7(M)$. A-priori the
number of independent harmonic sums of transcendentality 7 is quite
large, but we may use the
structure of the L{\"u}scher correction i.e. the fact that $S_1^2$
gets automatically factored out to significantly simplify the
analysis. Hence we may 
write $C_7(M)=S_1^2 C_5(M)$. We are thus left with harmonic sums of
transcendentality 5. Assuming that the index $-1$ does not appear we
are left with a basis of 41 harmonic sums. 

We then computed analytically the residues of the integrand at $q=iQ$
and summed up the resulting expression from $Q=1$ to $\infty$. We did
this for $M=1$ to $M=41$, where we included, similarly as in
\cite{KLRSV}, also the unphysical odd values of $M$. A justification
of the precise form of analytical continuation of the even $M$
integrands to odd $M$ is the agreement of the coefficients of
$\zeta(3)$ and $\zeta(5)$ and the fact that the remaining term was
just a rational number. The results from $M=1$ to $M=41$ fixed all
the coefficients of the 41 harmonic sums which turned out to be simple
integers. The result for the rational part obtained in this way is
\eq
\label{e.finlast}
C_7(M)=256 S_1^2 \left( -S_5+S_{-5}+2S_{4,1}-2S_{3,-2}+2S_{-2,-3}
-4S_{-2,-2,1} \right)
\eqx
As a check we then computed the wrapping correction from L{\"u}scher
formulas for $M=42$ and $M=44$ and got perfect agreement with the
above formulas (\ref{e.finfirst})-(\ref{e.finlast}).

In the remaining part of the paper we will first verify that
$\gm_8^{wrapping}(M)$ does not have any contribution of the order of
$\log M$ in the large $M$ limit so that the cusp anomalous dimension
remains unchanged. Then we will analyze the analytical continuation of
$\gm_8^{wrapping}(M)$ to $M=-1+\om$ and check compatibility with LO
and NLO BFKL. 

\section{Large $M$ asymptotics of $\gm_8^{wrapping}(M)$}

Wrapping corrections should not change the leading asymptotic behaviour of the
anomalous dimension of twist-2 operators. On the other hand the final
formula which we obtained is of the form  
$$
\gm_8^{wrapping}(M)=128 S_1^2(M) \times (\mbox{finite when } M\to\infty)
$$
where $S_1^2(M)$ scales as $\log^2(M)$ when $M\to \infty$. The only option for
the wrapping correction not to change the leading large $M$ asymptotics is
that the finite part vanishes at infinity. If this is the case then the first
wrapping contribution to the asymptotic behaviour may enter at the
order $\frac{\log^2(M)}{M}$ which is subleading comparing with $\log(M)$.   
 
Let us check if the leading expansion around $M=+\infty$ of our result
vanishes. The values at infinity of the nested harmonic sums can be expressed
in terms of multivariate zeta functions (Zagier-Euler sums) as was shown
in \cite{KV05}. These sums can be reexpressed in terms of ordinary Euler
sums. The relations between the former and the later can be found using
EZ-Face --- an on-line calculator for Euler sums \cite{EZFace}. 
The results relevant for us are   
\begin{eqnarray*}
S_{-2}(\infty) & =&-\frac{1}{2}\zeta(2)\\
S_{-5}(\infty) &=&-\frac{15}{16}\zeta(5)\\
S_{5}(\infty) &=&\zeta(5)\\
S_{4,1}(\infty) &=& -\zeta(2)\zeta(3)+3\zeta(5)\\
S_{3,-2}(\infty) &=&\frac{1}{4}\zeta(2)\zeta(3)-\frac{51}{32}\zeta(5)\\
S_{-2,-3}(\infty) &=&\frac{21}{8}\zeta(2)\zeta(3)-\frac{67}{16}\zeta(5)\\
S_{-2,-2,1}(\infty) &=&\frac{15}{16}\zeta(2)\zeta(3)-\frac{29}{32}\zeta(5)
\end{eqnarray*}

Plugging them into the coefficient of $S_1^2$ in the wrapping correction 
$$
-5\zeta(5)-4S_{-2}\zeta(3)-2S_{5}+2S_{-5}+4S_{4,1}-4S_{3,-2}+4S_{-2,-3}
-8S_{-2,-2,1}
$$
we obtain that it vanishes as we expected. It would be very
interesting to analyze the large $M$ limit in more detail along the
lines of \cite{bassokor,reciprocity}. 
 
\section{Analytical continuation of $\gm_8^{wrapping}(M)$}

The way how harmonic sums can be analytically continued was nicely
explained in \cite{KV05}. There is a pronounced difference between
the cases when all indexes are positive and when we have at least
one negative index. The former case can be described systematically,
while the latter one requires a case by case study. We relegate the
details of the analytical continuation of the needed harmonic sums
to Appendix C. Using the results there we can list the analytical
continuation of all harmonic sums of interest. Since we are concerned
with the singularity around $-1$ only up to the third order pole, it
is enough to expand $S_{1}$ to third order and keep only the singular
part of all other harmonic sums: 
\begin{eqnarray*}
S_{1}(-1+\om) & = & -\frac{1}{\om}+\om\zeta(2)-\om^{2}\zeta(3)+
\om^{3}\zeta(4)+\dots\\ 
S_{5}(-1+\om) & = & -\frac{1}{\om^{5}}+\dots\\
S_{4,1}(-1+\om) & = & -\frac{1}{\om^{4}}(\zeta(2)\om-\zeta(3)\om^{2}+
\zeta(4)\om^{3}+\dots)+\dots\\ 
S_{-5}(-1+\om) & = & \frac{1}{\om^{5}}+\dots\\
S_{3,-2}(-1+\om) & = & -\frac{1}{\om^{3}}(2\zeta(-2)-2\zeta(-3)\om+
3\zeta(-4)\om^{2}+\dots)+\dots\\ 
S_{-2,-3}(-1+\om) & = &
-\frac{1}{\om^{2}}(3\zeta(-4)\om-2\zeta(-3)+\dots)+ \dots\\
S_{-2,-2,1}(-1+\om) & = & -\frac{1}{\om^{2}}(-2S_{-2,1}(\infty)\\
 &  &
\;\qquad+\om(2S_{-3,1}(\infty)+S_{-2,2}(\infty)-\zeta(-2)\zeta(2)))+ 
\dots
\end{eqnarray*}
In the above expressions $\zeta$ of a negative argument denotes an
alternating version of the $\zeta$ function:
\eq
\zeta(-n)\equiv \sum_{k=1}^\infty \f{(-1)^k}{k^n}
\eqx
which is linked to the ordinary $\zeta$ function through the
well-known relation
\[
\zeta(-n)=(2^{1-n}-1)\zeta(n)
\]
The harmonic sums at infinity can be expressed in terms of Euler-Zagier
sums. We found the following relations useful:
$S_{-2,1}(\infty)=\frac{5}{6}\zeta(-3)$ 
and $2S_{-3,1}(\infty)+S_{-2,2}(\infty)=-\frac{37}{16}\zeta(4)$.
Some of them can be proven, but some we obtained using the program
EZ-Face \cite{EZFace}. To simplify the final form we also used that
$\zeta(2)^{2}=\frac{5}{2}\zeta(4)$. 
If we plug all these expressions into the wrapping correction 
\[ 
128S_{1}^{2}\left[-5\zeta(5)-4S_{-2}\zeta(3)-2S_{5}+2S_{-5}+
  4S_{4,1}-4S_{3,-2}+4S_{-2,-3}-8S_{-2,-2,1}\right] 
\]
we obtain the leading singularities around $M=-1+\om$ as 
\eq
\gm_8^{wrapping}(\om)\sim 
256\left(\frac{2}{\om^{7}}-\frac{8\zeta(2)}{\om^{5}}+
  \frac{9\zeta(3)}{\om^{4}}+\frac{59\zeta(4)}{4\om^{3}}+
  \oo{\f{1}{\om^2}} \right)
\eqx
which, when combined with the Bethe Ansatz result (\ref{e.bethean}) 
\eq
\gm_{8}^{Bethe}(\om) \sim 256\left( \f{-2}{\om^7}+ \f{0}{\om^6} +
\f{8\zeta(2)}{\om^5} -\f{13\zeta(3)}{\om^4} -\f{16 \zeta(4)}{\om^3}
+\oo{\f{1}{\om^2}} \right)
\eqx
agrees with LO and NLO BFKL expectations
\eq
\gm_{8}(\om) \sim -256\left( \f{4\zeta(3)}{\om^4} +
\f{\f{5}{4}\zeta(4)}{\om^3}  +\oo{\f{1}{\om^2}}  \right)
\eqx 

Finally we note that we can continue around any other negative integers
to compare with other predictions. In particular, it is known that there is an 
$\om^{-7}$ pole at even negative integers \cite{KLRSV}, whose coefficient is
fully reproduced just by the Bethe Ansatz answer. We will thus check that our wrapping correction does not give any contribution to this pole which we analyze at $M=-2+\om$. 
The terms which 
can contribute at this order are $S_{5}$ and $S_{-5}$. Since the
analytical continuations are $S_{\pm5}(M=-2+\om)=-\frac{1}{\om^{5}}$ there
is no contribution from wrapping at this order which is consistent
with the BA results.

\section{Conclusions}

In this paper, using integrability properties of the light-cone
quantized worldsheet QFT of the string in $AdS_5 \times S^5$, 
we have obtained the four-loop wrapping correction to the
anomalous dimension of twist-2 operators with spin $M$. As a first check 
we determined the large $M$ asymptotic behaviour of our expression and
concluded that it 
does not modify the Bethe Ansatz result for the cusp anomalous
dimension. 
In contrast, the wrapping
correction is essential for the correct behaviour of the twist-2
anomalous dimensions under analytic continuation to
$M=-1+\omega$. Indeed it exactly cancels all the higher order poles in
the Bethe Ansatz result and the remaining leading poles exactly agree
with LO and NLO BFKL predictions.

Let us note that despite the apparent complexity of the computation,
the string theory calculation performed in the present paper is still
much simpler than any corresponding gauge theory perturbative
computation. This suggests that one can use string theory methods of
the AdS/CFT correspondence as an efficient calculational tool even in
the weak coupling perturbative regime.

In fact, it would be interesting to analyze in detail the
interrelations between the perturbative and string theory computations
as it might give a hint on streamlining higher loop perturbative
methods.

Moreover it would be very interesting to use the complete formulas for
the 4-loop anomalous dimensions to understand better the underlying
theoretical structure both from the point of view of their asymptotic
properties along the lines of \cite{bassokor,reciprocity} and of the
analycity structure for higher negative integers as discussed in
\cite{KLRSV} and further aspects of the link with BFKL like in
\cite{GomezBFKL}.    

Apart from these physics issues, there still remain several aspects of
our derivation which may be improved. We 
have managed to do the exact calculation of the result for the terms
proportional to the $\zeta(3)$ and $\zeta(5)$ (see Appendix B) but the
derivation of the rational part was essentially based on the maximum
transcendentality conjecture. It would be very instructive to obtain this
result from first principles. 

Finally let us note that the exact form of the L{\"u}scher
corrections may be a precision test that has to be satisfied by any
yet-to-be constructed exact spectral equation (or more probably a set
of coupled nonlinear integral equations). Our computation in this
paper and the very nontrivial consistency requirement with the analytical
structure predicted from BFKL can be understood as a check of the form
of the L{\"u}scher corrections for the worldsheet QFT of the $AdS_5
\times S^5$ superstring.

\bigskip

\noindent{}{\bf Acknowledgments.} We thank Vitaly Velizhanin and the
authors of \cite{KLRSV} for sharing with us formula
(\ref{e.bethean}). This work has been supported in part by Polish
Ministry of Science and 
Information Technologies grant 1P03B04029 (2005-2008), RTN network
ENRAGE MRTN-CT-2004-005616, ToK grant COCOS MTKD-CT-2004-517186. 
ZB thanks the Jagellonian University for warm hospitality during the
time when this 
work was performed and for the Marie Curie ToK COCOS grant for 
the financal support. ZB was also supported by a Bolyai Scolarship and
OTKA 60040.

\appendix

\section{Evaluation of the integrand}

In this appendix we elaborate on the super-trace part of the integrand.
Since we expect the wrapping contribution to appear at order $g^{8}$
the super-trace part has to vanish at leading order. We check this
requirement in the first subsection and then make a systematical small
$g^{2}$ expansion of each term in the second one.

\subsection*{Zeroth order contribution}

Since the rapidities appear always in pairs $(u,-u)$ we found it
useful to combine their contributions. Using the explicit form of
these functions from \cite{BJ08} (see Appendix~D below) we evaluate
the expressions entering the evaluation of the supertrace
\[
SB0(q,u)SB0(q,-u)=1
\]
for any $g.$ To abbreviate the formulas we introduce the following
notation\[
f(g)=f_{0}(1+g^{2}\delta f+O(g^{4}))\]
 In this notation the single diagonal bosonic part is given by 
\[
SBQ(q,u)_{0}SBQ(q,-u)_{0}=\frac{(q+i(Q-1))^{2}-4u^{2}}{(q-i(Q-1))^{2}-
  4u^{2}}\]
While the diagonal fermionic part for $j=0,\dots,Q-1$ reads as \[
SF(q,u,j)_{0}SF(q,-u,j)_{0}=\frac{(q-i(Q-1)+2ij)^{2}-
4u^{2}}{(q-i(Q-1))^{2}-4u^{2}}\]
The main complication is in the matrix part. One can observe, however,
that evaluating the matrix part at $g=0$ the resulting matrix $SB(q,u,j)_{0}$
has eigenvectors $(\frac{4i(Q-j)}{(q+iQ)^{2}},1)$ and
  $(\frac{-4ij}{q^{2}+Q^{2}},1)$, 
which are independent of the $u_{k}$. Thus we can diagonalize all
of them simultaneously leading to 
\[
G\, SB(q,u,j)_{0}G^{-1}=\left(\begin{array}{cc}
\frac{i+2ij+q-iQ-2u}{i+q-iQ-2u} & 0\\
0 & \frac{2ij+q-i(1+Q)-2u}{i+q-iQ-2u}\frac{2u-i}{2u+i}\end{array}\right)\]
Let us denote the corresponding eigenvalues by $SB^{\pm}(q,u,j)_{0}.$
As a consequence \[
\mbox{Tr}\left(\prod_{k=1}^{M}SB(q,u_{k},j)\right)_{0}= 
\prod_{k=1}^{M}SB^{+}(q,u_{k},j)_{0}+\prod_{k=1}^{M}SB^{-}(q,u_{k},j)_{0}\]
where we have 
\[
SB^{\pm}(q,u,j)_{0}SB^{\pm}(q,-u,j)_{0}=\frac{(q-i(Q\mp1)+2ij)^{2}-
  4u^{2}}{(q-i(Q-1))^{2}-4u^{2}}
\] 
Now couple of observations are in order. We can see that \[
SB^{+}(q,u,0)_{0}SB^{+}(q,-u,0)_{0}=SB0(q,u)SB0(q,-u)=1\]
 and that \[
SBQ(q,u)_{0}SBQ(q,-u)_{0}=SB^{-}(q,u,Q)_{0}SB^{-}(q,-u,Q)_{0}\]
This means that we can incorporate the contribution of $SB0$ into
$SB^{+}$ by extending the summation over $j$ in (\ref{eq:STr}) from
$j=1$ to $j=0$. 
In a similar manner, the summation of the contributions of $SB^{-}$ when
extended naively to $j=Q$ turns out to automatically incorporate the
contribution 
of $SBQ$. 
We can further realize that 
\[
SB^{+}(q,u,j)_{0}SB^{+}(q,-u,j)_{0}=SB^{-}(q,u,j+1)_{0}SB^{-}(q,-u,j+1)_{0}
\]
Thus the two bosonic eigenvalues contribute in the same way. This means that
at leading order the super-trace part reads as 
\[
\mbox{STr}\Bigl(\prod_{k=1}^{M}S_{Qmatrix}^{sl(2)}(q,u_{k})\Bigr)_{0}=
-2\sum_{j=0}^{Q-1}\prod_{k=1}^{M}SF(q,u_{k},j)_{0}+
2\sum_{j=0}^{Q-1}\prod_{k=1}^{M}SB^{+}(q,u_{k},j)_{0}
\]
But finally we can observe that the bosonic and the fermionic contributions
are the same for each $j$:
\[
SB^{+}(q,u,j)_{0}SB^{+}(q,-u,j)_{0}=SF(q,u,j)_{0}SF(q,-u,j)_{0}
\]
so the super-trace part vanishes at leading order and consequently
the wrapping contribution also vanishes at order $g^{4}.$

\subsection*{First order contribution}

We expand now each function to the order $g^{2}$ as follows: The
expansion of $SB0$ is trivial, the correction vanishes. The other
diagonal bosonic contribution reads as \[
\prod_{k=1}^{M}SBQ(q,u_{k})=\Bigl(\prod_{k=1}^{M}SBQ(q,u_{k})_{0}\Bigr)\left[1+g^{2}\sum_{k=1}^{M}\delta SBQ(q,u_{k})\right]\]
The fermionic is given by\[
\prod_{k=1}^{M}SF(q,u_{k},j)=\Bigl(\prod_{k=1}^{M}SF(q,u_{k},j)_{0}\Bigr)\left[1+g^{2}\sum_{k=1}^{M}\delta SF(q,u_{k},j)\right]\]
In the matrix case we use the same matrices $G$ we used to diagonalize
$SB(q,u,j)_{0}$ and bring $G\, SB(q,u,j)G^{-1}$ into the form\[
\left(\begin{array}{cc}
SB^{+}(q,u,j)_{0}(1+g^{2}\delta SB^{+}(q,u,j)) & O(g^{2})\\
O(g^{2}) & SB^{-}(q,u,j)_{0}(1+g^{2}\delta SB^{-}(q,u,j))\end{array}\right)\]
The advantage of this form is that \begin{eqnarray*}
\mbox{Tr}\left(\prod_{k=1}^{M}SB(q,u_{k},j)\right) & = & \Bigl(\prod_{k=1}^{M}SB^{+}(q,u_{k},j)_{0}\Bigr)\left[1+g^{2}\sum_{k=1}^{M}\delta SB^{+}(q,u_{k},j)\right]+\\
 &  & \Bigl(\prod_{k=1}^{M}SB^{-}(q,u_{k},j)_{0}\Bigr)\left[1+g^{2}\sum_{k=1}^{M}\delta SB^{-}(q,u_{k},j)\right]\end{eqnarray*}
We have checked again by explicit calculation that \[
\delta SB^{+}(q,u,0)+\delta SB^{+}(q,-u,0)=0\]
and that \[
\delta SBQ(q,u)+\delta SBQ(q,-u)=\delta SB^{-}(q,u,Q)+\delta SB^{-}(q,-u,Q)\]
Thus both bosonic summations can be extended as before. Shifting the
summation in the $SB^{-}$ case we can bring the whole leading correction
into the form:\begin{eqnarray*}
\mbox{STr}\Bigl(\prod_{k=1}^{M}S_{Qmatrix}^{sl(2)}(q,u_{k})\Bigr) & = & g^{2}\sum_{j=0}^{Q-1}\Bigl(\prod_{k=1}^{M}SB^{+}(q,u_{k},j)_{0}\Bigr)\sum_{k=1}^{M}\delta SBF(q,u_{k},j)\end{eqnarray*}
where \[
\delta SBF(q,u_{k},j)=\delta SB^{+}(q,u_{k},j)+\delta SB^{-}(q,u_{k},j+1)-2\delta SF(q,u_{k},j)\]
We could explicitly calculate this quantity, which turned out to be
\[
\delta SBF(q,u_{k},j)=\frac{16}{1+4u_{k}^{2}}\left[\frac{1}{2j-iq-Q}-\frac{1}{2(j+1)-iq-Q}\right]\]

\section{Exact calculation of the coefficient of $\zeta(3)$ and $\zeta(5)$}

Having taken the residue at $q=iQ$ we will encounter derivatives
of the Baxter's Q functions (denoted here by $P_M(.)$ to avoid confusion
with $Q$) around four different points, namely around
$\pm\frac{i}{2}$ and $\pm\frac{i}{2}+iQ$. In the following calculation
it is more convenient to use the function with a shifted argument
\[
U(\pm Q)=P_M(\pm\frac{i}{2}+iQ)
\]
It has the following expansion around $Q=0$ : 
\[
U(Q)=1+2S_{1}\, Q+2(S_{1}^{2}+S_{-2})\, Q^{2}+\dots
\]
After taking the residue of the integrand we have a sum of rational
functions of $Q$ where the denominators are products of powers
of $Q$ and $U(Q)U(-Q)$. We can make a partial fraction expansion
of this result and separate the terms having denominator $Q^{n}$:\[
\frac{A_{5}}{Q^{5}}+\frac{A_{4}}{Q^{4}}+\frac{A_{3}}{Q^{3}}+
\frac{A_{2}}{Q^{2}}+\frac{A_{1}}{Q}+\sum_{i,j=1}^{i+j\leq5} 
\frac{A_{ij}(Q)}{U(Q)^{i}U(-Q)^{j}}\]
Explicit calculation showed that the quantities of interest read as
\[
A_{5}=10\qquad;\quad A_{3}=U^{\prime\prime}(0)-4U^{\prime}(0)^{2}+ 
12(\partial_{Q}\widetilde{T}(q,Q))\vert_{q=iQ=0}\]
In writing $A_{3}$ into the form above we used $U(0)=1$ and the
previously calculated values of $A_{5}$ and
$A_{4}=12(\widetilde{T}(0,0)+U^{\prime}(0))=0$. 
The other coefficients $A_{2}$ and $A_{1}$ and especially
the polynomials $A_{ij}(Q)$ are very cumbersome to compute. Using
the expansion of $U(Q)$ in terms of harmonic sums together with 
$(\partial_{Q}\widetilde{T}(q,Q))\vert_{q=iQ=0}=S_{1}^{2}-S_{-2}$
we obtain the form of the wrapping correction 
\[
\Delta E=-128g^{8}\,
S_{1}^{2}\,\left[5\zeta(5)+4S_{-2}\zeta(3)+\dots\right]
\]
where the ellipsis denotes a rational number which can be expanded
in terms of harmonic sums and which we determine in section~6. 

We note that using perturbative methods in the $\mathcal{N}=4$
super Yang-Mills theory the coefficient of $\zeta(5)$ was
calculated in \cite{Vel08}. It is exactly reproduced by our
computation. The conjecture for the coefficient of $\zeta(3)$ given in
\cite{Vel08} disagrees, however, with our result. 

\section{Analytical continuation of harmonic sums}

There is a pronounced difference in the analytical continuation of
harmonic sums between the cases when all indexes are positive and
when we have at least one negative index. The former case can be described
systematically, while the latter one requires a case by case study.

\subsection*{Analytical continuation with only positive indices}

The analytical continuation of harmonic sums with all indices positive
can be done inductively. One starts with the simplest one $S_{a}(n)$
and uses the general strategy to move the variable $n$ from the upper
bound of the sum to the summand as \[
S_{a}(n)=\left(\sum_{j=1}^{\infty}-\sum_{j=n+1}^{\infty}\right)\frac{1}{j^{a}}=S_{a}(\infty)-\sum_{k=1}^{\infty}\frac{1}{(k+n)^{a}}\]
Since we are interested in the analytical continuation around $-1$
we explicitly separate the singular and regular pieces as \[
S_{a}(-1+x)=-\frac{1}{x^{a}}+S_{a}(\infty)-d_{a}(x)\]
where we found it useful to introduce the function\[
d_{\pm a}(x)=\sum_{k=1}^{\infty}\frac{(\pm1)^{k}}{(k+x)^{a}}\]
This function is regular around $x=0$ and has the expansion\[
d_{a}(x)=\zeta(a)-xa\zeta(a+1)+x^{2}{a+1 \choose 2}\zeta(a+2)+\dots+(-1)^{n}x^{n}{a+n-1 \choose n}\zeta(a+n)+\dots\]
Suppose now that we have already analytically continued $S_{b,\dots,c}(n)$
and then we want to continue analitically $S_{a,b,\dots,c}(n)$. Using
the previous strategy we can write\[
S_{a,b,\dots,c}(n)=\left(\sum_{j=1}^{\infty}-\sum_{j=n+1}^{\infty}\right)\frac{1}{j^{a}}S_{b,\dots,c}(j)=S_{a,b,\dots,c}(\infty)-\sum_{k=1}^{\infty}\frac{1}{(k+n)^{a}}S_{b,\dots,c}(k+n)\]
Let us specify this formula for the case of interest, namely to $S_{a,b}(n)$
around $n=-1.$ Focusing only on the singular part, which comes from
the $k=1$ term, we have \[
S_{a,b}(-1+x)=-\frac{1}{x^{a}}\left[S_{b}(x)\right]+\mbox{reg}\]
 where by {}``reg'' we mean terms regular for $x\to0$ and \[
S_{b}(x)=S_{b}(\infty)-d_{b}(x)=xb\zeta(b+1)-x^{2}{b+1 \choose 2}\zeta(b+2)+\dots\]
This is all we need for the analytical continuation for positive indexes.

\subsection*{Analytical continuation with at least one negative index }

In the case when we have at least one negative index in the harmonic
sum we have to be careful. The main problem is that the harmonic sum
determines two different analytical functions, depending on whether
we continue from even or from odd values. Since in our problem we
continue from even ones we have to extend our original harmonic sum
to odd values first, then to do the analytical continuation. Since
the way how one can do this extension for the functions we need is
thoroughly explained in \cite{KV05} we merely cite the result. For
instance, in the simplest case when we have just one index the function
\[
\bar{S}_{-a}^{+}(n)=(-1)^{n}S_{-a}(n)+(1-(-1)^{n})S_{-a}(\infty)\]
is the same as $S_{-a}(n)$ for even values and is extended to odd
values in such a way that it can be described by one function. In
calculating the analytical continuation around $-1$ we use the method
of moving $n$ from upper bound to the summand and obtain\[
\bar{S}_{-a}^{+}(-1+x)=\frac{1}{x^{a}}+S_{-a}(\infty)+d_{-a}(x)\]
where the function $d_{-a}(x)$ is regular around zero and have the
expansion: \[
d_{-a}(x)=\zeta(-a)-xa\zeta(-a-1)+x^{2}{a+1 \choose 2}\zeta(-a-2)+\dots\]
Similarly one can define the extension of the harmonic sums with the
first index being negative to be \[
\bar{S}_{-a,b,\dots,c}^{+}(n)=(-1)^{n}S_{-a,b,\dots,}(n)+(1-(-1)^{n})S_{-a,b,\dots c}(\infty)\]
as the proper continuation from even values to odd ones. Its analytical
continuation around $-1$ has the singular part \[
\bar{S}_{-a,b,\dots,c}^{+}(-1+x)=\frac{1}{x^{a}}\left[S_{b,\dots,c}(x)\right]+\mbox{reg}\]
The analytical continuation of the function $S_{a,-b}(n)$ is more
tricky and one has to use the functional relations the harmonic sums
satisfy to define\[
\bar{S}_{a,-b}^{+}(n)=(-1)^{n}S_{a,-b}(n)+(1-(-1)^{n})(S_{a,-b}(\infty)-S_{b}(\infty)(S_{a}(\infty)-S_{a}(n))\]
Thus the singular part around $-1$ reads as \[
\bar{S}_{a,-b}^{+}(-1+x)=-\frac{1}{x^{a}}(2S_{-b}(\infty)-\bar{S}_{-b}^{+}(x))=-\frac{1}{x^{a}}(S_{-b}(\infty)+d_{-b}(x))\]
We also need the analytical continuation of $S_{-2,-2,1}(n)$ so we
need \[
\bar{S}_{-a,-b,c}^{+}(n)=\sum_{j=1}^{n}\frac{1}{j^{a}}\bar{S}_{-b,c}^{+}(j)+S_{-b,c}(\infty)(\bar{S}_{a}^{+}(n)-S_{a}(n))\]
Its analytical continuation has a singular part: \[
\bar{S}_{-a,-b,c}^{+}(-1+x)=-\frac{1}{x^{a}}\bar{S}_{-b,c}^{+}(x)+S_{-b,c}(\infty)(\frac{1}{x^{a}}-\frac{-1}{x^{a}})\]
in which we have \begin{eqnarray*}
\bar{S}_{-b,c}^{+}(x) & = & S_{-b,c}(\infty)-\sum_{k=1}^{\infty}\frac{(-1)^{k}}{(k+x)^{b}}S_{c}(k+x)\\
 & = & x(bS_{-b-1,c}(\infty)+cS_{-b,c+1}(\infty)-cS_{-b}(\infty)S_{c+1}(\infty))+\dots\end{eqnarray*}

\section{Explicit S-matrix coefficients}

In this appendix we collect the explicit expressions for the matrix
part of the $Q-1$ scattering matrix derived in \cite{BJ08}. 
We choose the normalization
as \[
SB0(q,u)=a_{1}^{1}(q,u)=1\]
The bosonic matrix part reads as

\[
SB(q,u,j)=\left(\begin{array}{cc}
\frac{Q+1-j}{Q+1}a_{1}^{1}(q,u)+\frac{j}{2}a_{2}^{2}(q,u) & \frac{j(Q-j)}{Q-1}a_{4}^{2}(q,u)\\
\frac{1}{Q-1}a_{2}^{4}(q,u) & \frac{2(Q-j)}{(Q-1)^{2}}a_{4}^{4}(q,u)
+\frac{j-1}{2}a_{10}^{10}(q,u)\end{array}\right)\]
where $j=1,\dots ,Q-1$. The last bosonic part is simply

\[
SQ(q,u)=\frac{1}{Q+1}a_{1}^{1}(q,u)+\frac{Q}{2}a_{2}^{2}(q,u)\]
while the fermionic part is given by

\[
SF(q,u,j)=\frac{Q-j}{Q^{2}}a_{6}^{6}(q,u)+\frac{j}{2}a_{7}^{7}(q,u)\]
The various matrix elements read in terms of $x^{\pm}(u)$ and $z^{\pm}(q)$
as \[
a_{6}^{6}=Q\frac{x^{-}-z^{-}}{x^{+}-z^{-}}\sqrt{\frac{x^{+}}{x^{-}}}\qquad;
\quad a_{7}^{7}=\frac{2}{Q}\frac{z^{-}(x^{-}-z^{+})(1-x^{-}z^{+})}{z^{+}
(x^{+}-z^{-})(1-x^{-}z^{-})}\sqrt{\frac{x^{+}}{x^{-}}}\]
\[
a_{10}^{10}=\frac{2}{Q-1}\frac{z^{-}(x^{-}-z^{+})(1-x^{+}z^{+})}
{z^{+}(x^{+}-z^{-})(1-x^{-}z^{-})}\]
\[
a_{4}^{2}=-i\frac{Q-1}{Q}\frac{z^{-}(x^{-}-x^{+})}{z^{+}(x^{+}-z^{-})(1-x^{-}z^{-})}
\qquad;\quad a_{2}^{4}=i\frac{Q-1}{Q}\frac{(z^{-}-z^{+})^{2}(x^{-}-x^{+})}{(x^{+}-z^{-})(1-x^{-}z^{-})}\]
\begin{eqnarray*}
a_{2}^{2}=-\frac{1}{Q(1+Q)}\frac{1}{z^{+}(x^{+}-z^{-})(1-x^{-}z^{-})}
\left[2z^{-}z^{+}(Q+x^{-}z^{-}-(1+Q)x^{-}z^{+})\right.\\
\left.+2x^{+}(z^{+}+z^{-}(-1+Q(-1+x^{-}z^{+})))\right]\end{eqnarray*}
\begin{eqnarray*}
a_{4}^{4}=-\frac{(Q-1)}{2Qx^{+}x^{-}(x^{+}-z^{-})(1-x^{-}z^{-})}
\left[x^{-}(Q(x^{-})^{2}x^{+}z^{-}-x^{-}(x^{+}+z^{-})\right.\\
\left.+x^{+}z^{-}(2-x^{+}z^{-}))-(x^{-}-x^{+})x^{+}z^{-}(z^{-}-z^{+}))\right]\end{eqnarray*}

\end{document}